\documentclass[aps,prm,twocolumn,showpacs,amsmath,amssymb,superscriptaddress,citeautoscript]{revtex4-2}

\usepackage{graphicx}
\usepackage{natbib}
\usepackage{dcolumn}
\usepackage{bm}
\usepackage{color}

\newcommand{\MN}[2]{{\color{black}#2}}                

\begin{document}

\title{Pressure tuning of the anomalous Hall effect in the chiral antiferromagnet Mn$_{3}$Ge}

\author{R.~D.~dos Reis}
\affiliation{Max Planck Institute for Chemical Physics of Solids, N\"{o}thnitzer Str.\ 40, 01187 Dresden, Germany}
\affiliation{Brazilian Synchrotron Light Laboratory (LNLS), Brazilian Center for Research in Energy and Materials (CNPEM), Campinas, Sao Paulo, Brazil}

\author{M.~Ghorbani Zavareh}
\email{Present address: HC DI CT R\&D CTC SA, Siemens Healthcare GmbH, Siemensstr.\ 3, 91301 Forchheim, Germany}
\affiliation{Max Planck Institute for Chemical Physics of Solids, N\"{o}thnitzer Str.\ 40, 01187 Dresden, Germany}

\author{M.~O.~Ajeesh}
\affiliation{Max Planck Institute for Chemical Physics of Solids, N\"{o}thnitzer Str.\ 40, 01187 Dresden, Germany}

\author{L.~O.~Kutelak}
\affiliation{Brazilian Synchrotron Light Laboratory (LNLS), Brazilian Center for Research in Energy and Materials (CNPEM), Campinas, Sao Paulo, Brazil}

\author{A.~S.~Sukhanov}
\affiliation{Max Planck Institute for Chemical Physics of Solids, N\"{o}thnitzer Str.\ 40, 01187 Dresden, Germany}

\author{Sanjay Singh}
\affiliation{Max Planck Institute for Chemical Physics of Solids, N\"{o}thnitzer Str.\ 40, 01187 Dresden, Germany}
\affiliation{School of Materials Science and Technology, Indian Institute of Technology (BHU), Varanasi-221005, India}

\author{\MN{}{J.~Noky}}
\affiliation{Max Planck Institute for Chemical Physics of Solids, N\"{o}thnitzer Str.\ 40, 01187 Dresden, Germany}

\author{\MN{}{Y.~Sun}}
\affiliation{Max Planck Institute for Chemical Physics of Solids, N\"{o}thnitzer Str.\ 40, 01187 Dresden, Germany}

\author{J.~E.~Fischer}
\affiliation{Max Planck Institute for Chemical Physics of Solids, N\"{o}thnitzer Str.\ 40, 01187 Dresden, Germany}

\author{K.~Manna}
\affiliation{Max Planck Institute for Chemical Physics of Solids, N\"{o}thnitzer Str.\ 40, 01187 Dresden, Germany}

\author{C.~Felser}
\affiliation{Max Planck Institute for Chemical Physics of Solids, N\"{o}thnitzer Str.\ 40, 01187 Dresden, Germany}

\author{M.~Nicklas}
\email{michael.nicklas@cpfs.mpg.de}
\affiliation{Max Planck Institute for Chemical Physics of Solids, N\"{o}thnitzer Str.\ 40, 01187 Dresden, Germany}

\begin{abstract}

We report on the pressure evolution of the giant anomalous Hall effect (AHE) in the chiral antiferromagnet Mn$_{3}$Ge. The AHE originating from the non-vanishing Berry curvature in Mn$_{3}$Ge can be continuously tuned by application of hydrostatic pressure. At room temperature,
the Hall signal changes sign as a function of pressure and vanishes completely at $p=1.53$~GPa. Even though the Hall conductivity changes sign upon increasing pressure, the room-temperature saturation value of $23~{\rm \Omega^{-1}cm^{-1}}$ at 2.85~GPa is remarkably high and comparable to the saturation value at ambient pressure of about $40~{\rm \Omega^{-1}cm^{-1}}$. The change in the Hall conductivity can be directly linked to a gradual change of the size of the in-plane components of the Mn moments in the non-collinear triangular magnetic structure.
Our findings, therefore, provide a route for tuning of the AHE in the chiral antiferromagnetic Mn$_{3}$Ge.

\end{abstract}

\date{\today}

\maketitle


The search for effective control and manipulation of spin degrees of freedom in solid-state systems, has attracted huge attention in the past three decades due the potential for applications \cite{Novoselov,Chappert_nmat207,Kent_Nat_nano2015,Waldrop_Nature,Enke_natphys2018,Faleev_PhysRevMaterials}. So far most spintronic devices are based on ferromagnetic materials. Only recently antiferromagnetic materials came into focus due to many new phenomena, such as the large anomalous Hall effect (AHE), spin Hall magnetoresistance, skyrmions, and the spin Seebeck effect, enriching the microcosmic physics system and encouraging vigorous development of the new field of antiferromagnetic spintronics \cite{Jungwirth_Nature_Phys18,Nayak_2016,Wadley587,Xiao2012,Manna2018}. This field aims to complement or replace ferromagnets in the active components of spintronic devices. Compared to ferromagnets, antiferromagnetic materials resist perturbation well, producing no stray fields that perturb the neighboring cells. The latter removes an obstacle for high-density memory integration. Furthermore, they demonstrate ultrafast dynamics and generate large magneto-transport effects. In that sense, antiferromagnetic materials have become increasingly important and exhibit various promising applications including non-volatile memory and magnetic field probes. However, although several schemes have been theoretically proposed to achieve full control of an antiferromagnetic structure, the experimental demonstration is still challenging and remains elusive \cite{Marrows558,Park_B_NatureMaterials_2011,Ohuchi_2018}. Therefore, finding ways to efficiently manipulate the magnetic states in antiferromagnets is the key to the further development.

The  Heusler compound Mn$_{3}$Ge is known to crystallize in both hexagonal and tetragonal structures \cite{Zhang_2013}. While the tetragonal material presents ferrimagnetic ordering below the Curie temperature of 710~K, the hexagonal exhibits a triangular antiferromagnetic structure with an ordering temperature of $365 - 400$~K \cite{Ohoyama_JPSJP}. In the hexagonal structure, the  unit cell of Mn$_{3}$Ge consists of two layers of Mn stacked along the \MN{}{$[0001]$}-axis. In each layer, the Mn atoms form a Kagom\'{e} lattice, with Ge sitting at the center of a hexagon \cite{Nayak_2016,Ohoyama_JPSJP,Qian_2014,Qian_2014}. Previous neutron diffraction experiments characterize the spin arrangement as a noncollinear triangular antiferromagnetic lattice of the Mn spins in which neighboring moments are aligned in the \MN{}{$(0001)$}-plane at an angle of $120^{\circ}$ \cite{Nagamya1979,NAGAMIYA1982,Sukhanov,soh2020ground}. In this configuration the spin triangle is free to rotate in an external magnetic field and, due the geometrical frustration of the Mn moments, a weak ferromagnetism is observed along the $[0001]$-axis.

\begin{figure}[t!]
\begin{center}
  \includegraphics[clip,width=0.98\columnwidth]{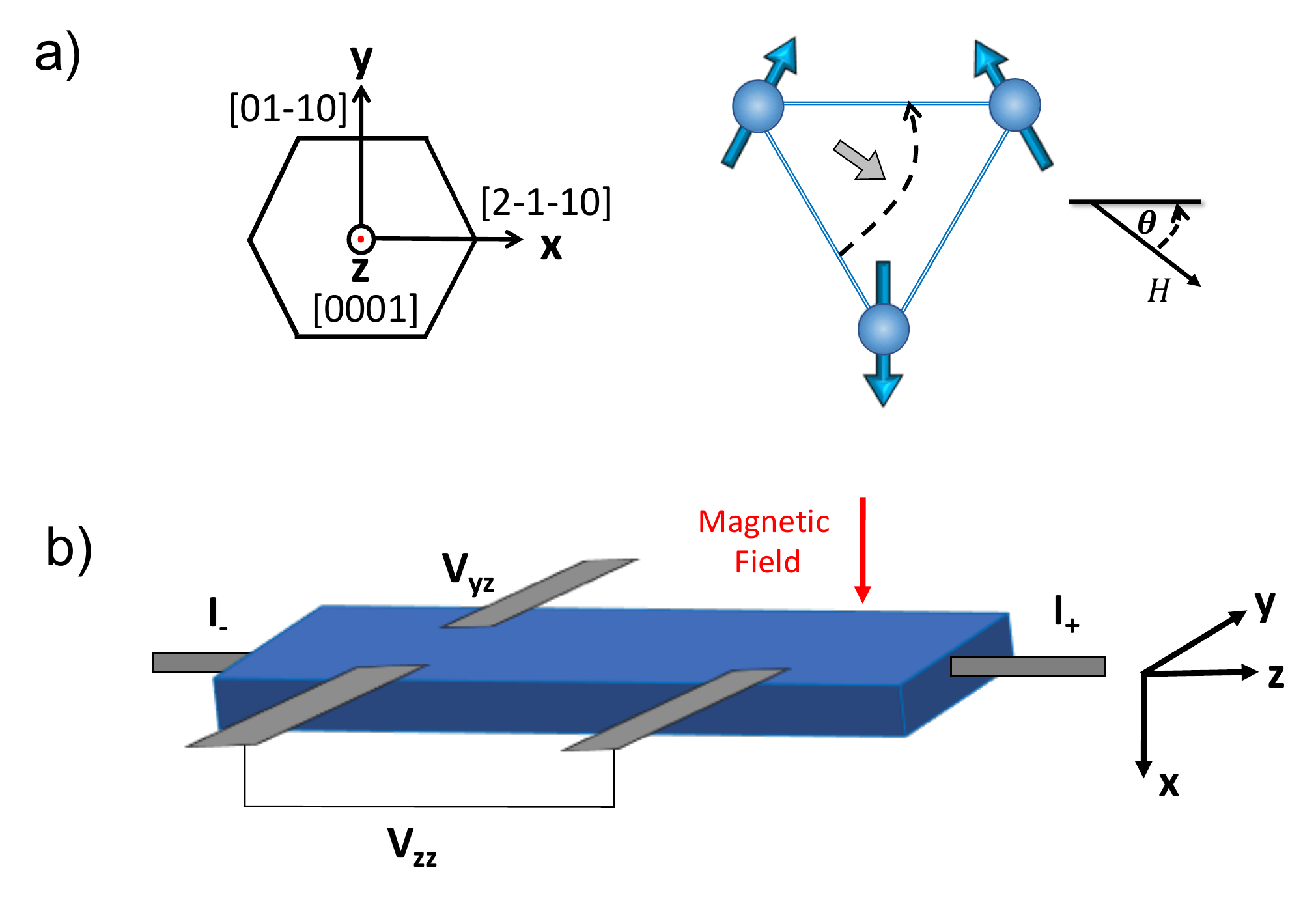}
  \caption{ \MN{}{(a)} Illustration of the rotation of the triangular spin structure upon application of an in-plane magnetic field. The direction of in-plane net magnetic moment is represented by the gray arrow. \MN{}{(b) Schematic drawing of the sample device for the electrical transport measurements used inside the pressure cell.}
  }\label{fig1}
  \end{center}
\end{figure}

Theoretical predictions demonstrate that this noncollinear triangular antiferromagnetic structure, common to all Mn$_{3}X$ ($X={\rm Ge}$, Sn, Ga, Ir, Rh, and Pt) compounds, gives origin to a nonvanishing Berry curvature which was predicted to lead to a large AHE \cite{Zhang_2013,Kubler_2014,Chen_PhysRevLett2014,Shindou_PhysRevLett.2001,Nakatsuji_2015,PhysRevB.95.075128,Zhang_2018}. Indeed, Nayak \textit{et al.}\ \cite{Nayak_2016} reported a large  AHE effect comparable to that observed for ferromagnetic metals even at room temperature \cite{Taguchi2573,Enke_natphys2018,Ghimire_2018}. Supported by the angular dependence of the AHE, they demonstrated that the small residual in-plane magnetic moment plays no role in the observed effect except to control the chirality of the triangular spin structure [see Fig.\ \ref{fig1}a]. Thus, they concluded that the large anomalous Hall effect in Mn$_{3}$Ge originates from a nonvanishing Berry curvature that arises from the chiral spin structure. \textit{Ab initio} band structure calculations indicate multiple Weyl points in the bulk band structures of the chiral antiferromagnetic compound Mn$_{3}$Ge with the positions and chirality of Weyl points in accordance with the symmetry of the magnetic lattice \cite{Yang_2017, Kubler_2017}. The discovery of Weyl points verifies that the large anomalous Hall conductivity observed is indeed connected to the electronic structure of the material \cite{Manna2018,Wuttke2019}.

The exotic transport properties make hexagonal Mn$_{3}$Ge not only interesting for fundamental research, but also a promising material for novel applications. Mn$_{3}$Ge belongs to the family of Heusler compounds which is known for their high structural tunability by chemical substitution and external pressure\cite{Casper_2012,Graf,Manna2018}. Although chemical substitution is useful to study the relation between lattice parameters and other physical properties, it also leads to an increase in disorder, which in this case can thwart the exotic transport properties. The use of external pressure, on the other hand, is a powerful approach to continuously tune the physical properties of intermetallic compounds and appears to bethe prime choice for controlling the noncollinear triangular antiferromagnetic structure of hexagonal Mn$_{3}$Ge.

In this Rapid Communication, we demonstrate that the AHE in hexagonal Mn$_{3}$Ge can be \MN{continuously}{} controlled by hydrostatic pressure. Combining our electrical-transport data with previous powder neutron diffraction data allows us to correlate the changes in the AHE amplitude with the gradual change in the magnetic structure. This result demonstrates a way to continuously tune the AHE at room temperature by hydrostatic pressure.


High-quality single crystals of hexagonal Mn$_{3}$Ge were grown using the Bridgman-Stockbarger technique. More details on the sample preparation and ambient pressure characterization can be found elsewhere \cite{Nayak_2016}. The electrical-transport experiments were performed in magnetic fields up to $B=9$~T in a Quantum Design Physical Property Measurement System (PPMS-9T) at temperatures from room temperature down to $T=5$~K. Electrical transport data were recorded in 5-point geometry, where the electrical contacts were spot welded to the sample using $25~\mu$m platinum wire. The electrical current was applied parallel to the crystallographic \MN{}{$[0001]$}-axis and the magnetic field parallel to the \MN{}{[2--1--10]}-axis. A schematic plot of the electrical-transport experiment setup is displayed in Fig.\ \ref{fig1}(b). Hydrostatic pressure was generated using a clamp-type pressure cell with silicon oil as the pressure transmitting medium. The pressure inside the cell was determined by the shift of the superconducting critical temperature of a piece of lead placed next to the sample. The narrow width of the superconducting transition at all pressures confirmed the good hydrostatic conditions in the sample chamber.


\begin{figure}[t!]
\begin{center}
  \includegraphics[clip,width=0.98\columnwidth]{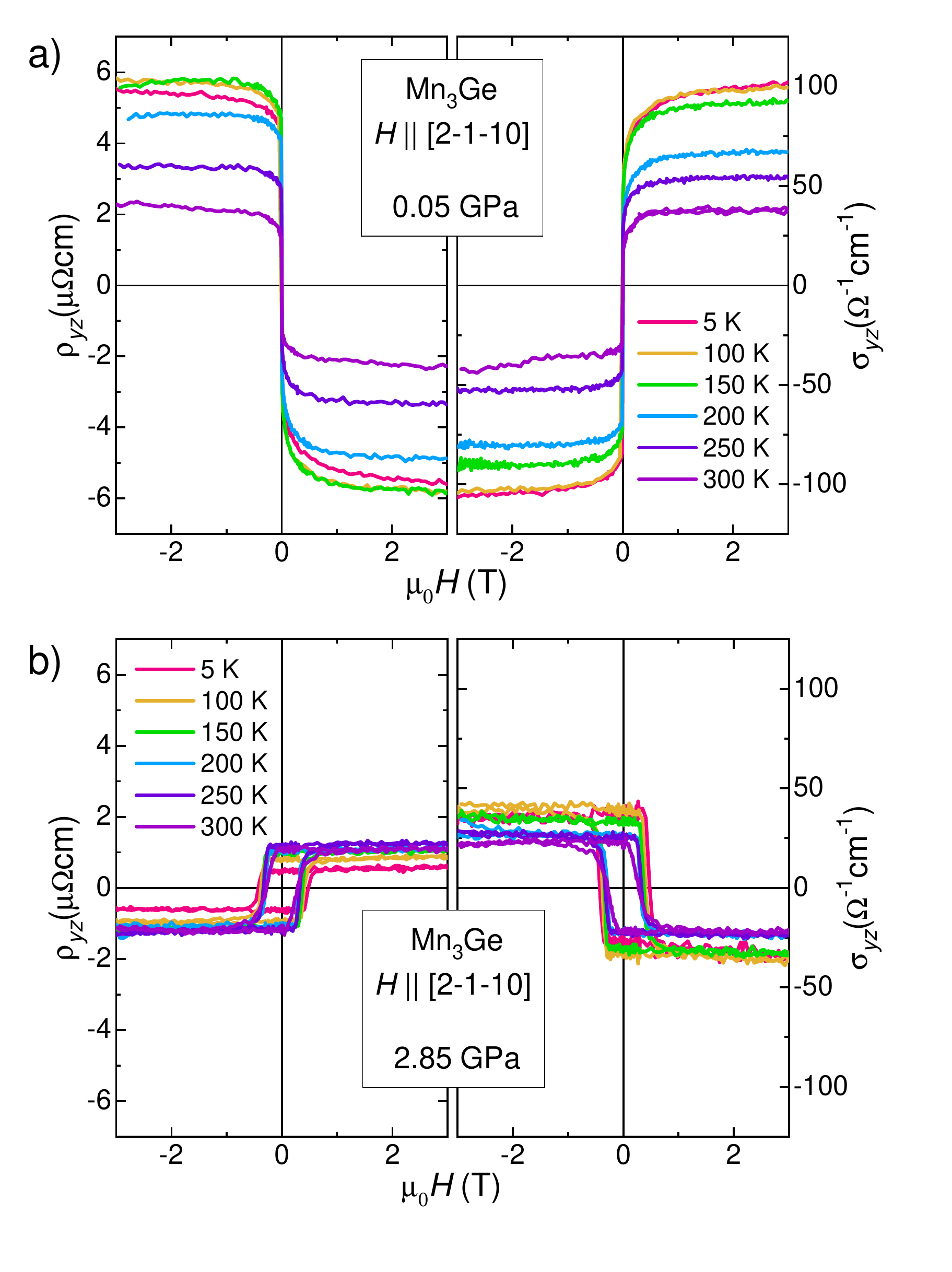}
  \caption{ (a) and (b) Hall resistivity \MN{}{($\rho_{yz}$)}  and Hall conductivity \MN{}{($\sigma_{yz}$)} as a function of magnetic field ($H$) at representative temperatures between 5 and 300~K for pressures of 0.05 and 2.85~GPa, respectively.
  }\label{fig2}
  \end{center}
\end{figure}

In order to probe the effect of hydrostatic pressure on the AHE in Mn$_{3}$Ge we carried out detailed electrical-transport measurements as a function of magnetic field for temperatures between 5 and 300~K at pressures up to 2.85~GPa. For that, the sample was cooled in zero magnetic field down to 5~K, and Hall and magnetoresistance were measured in a field loop from $-9$ to 9~T and back to $-9$~T again. Consecutively, the next temperatures were approached and loops between $-3$ and 3~T recorded.
From the Hall and transversal magnetoresistance we calculated the Hall conductivity using the expression \MN{}{$\sigma_{H}=-\rho_{yz}/(\rho_{yz}^{2}+\rho_{zz}^{2})$}, where \MN{}{$\rho_{yz}$} and \MN{}{$\rho_{zz}$} are the Hall and transversal resistivity, respectively. We note that the total (observed) Hall conductivity \MN{}{$\sigma_{yz}$} is the sum of the conventional Hall and anomalous Hall conductivities. Figure \ref{fig2} provides an overview of the pressure effects on the AHE of Mn$_{3}$Ge. At ambient pressure, see Fig.\  \ref{fig2}(a), both \MN{}{$\rho_{yz}(H)$} and \MN{}{$\sigma_{yz}(H)$} abruptly saturate in small magnetic fields of only about $\pm0.1$~T. The saturation value is large, $|\MN{}{\sigma^{sat}_{yz}}|\approx 100~{\rm \Omega^{-1}cm^{-1}}$ at 5~K and $|\MN{}{\sigma^{sat}_{yz}}|\approx 40~{\rm \Omega^{-1}cm^{-1}}$ at 300~K. This is in good agreement with previous studies \cite{Nayak_2016,Nakatsuji_2015}. At first glance, the large AHE at ambient pressure appears to be due to an antiferromagnetic ordering with a noncollinear spin arrangement that leads to a nonvanishing Berry phase in Mn$_{3}$Ge. At $p=2.85$~GPa, on another side, the highest pressure in our experiment, the amplitude of the Hall signal is inverted with respect to the magnetic field. The data, presented in Fig.\  \ref{fig2}(b), become almost temperature independent. There is only a small reduction in $|\rho^{sat}_{yz}|$ and a small increase in $|\sigma^{sat}_{yz}|$ upon decreasing temperature, respectively. The magnitude of $|\MN{}{\rho^{sat}_{yz}}|$ and $|\MN{}{\sigma^{sat}_{yz}}|$ is only slightly reduced compared with ambient pressure at 300~K, while there is a much stronger reduction at lower temperatures.
In addition, we observe a small hysteresis at 2.85~GPa in the \MN{}{$\rho_{yz}(H)$} and \MN{}{$\sigma_{yz}(H)$} loops whose origin is discussed below.

\begin{figure}[tb!]
	\begin{center}
		\includegraphics[width=0.98\columnwidth]{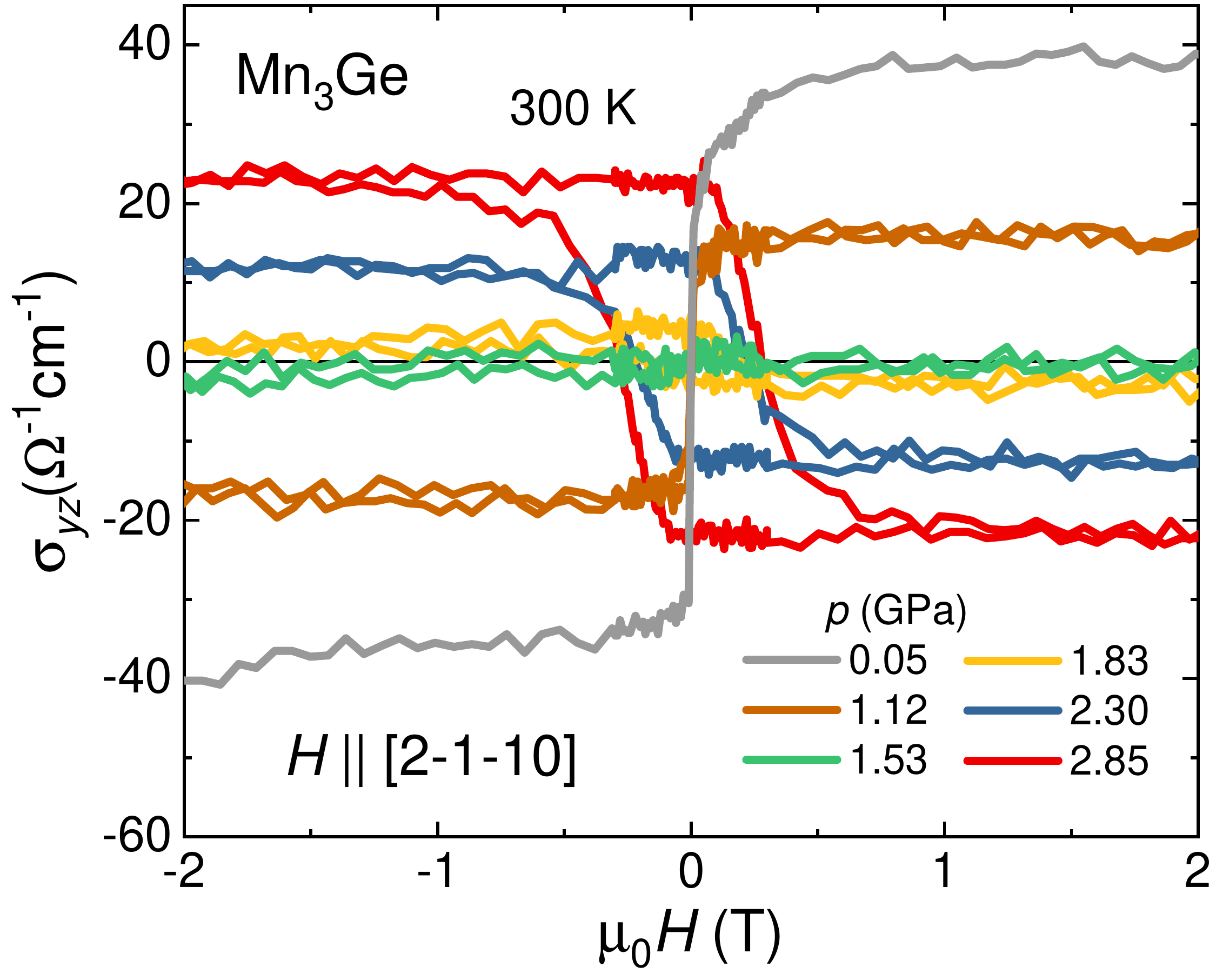}
		\caption{Field dependence Hall conductivity for Mn$_{3}$Ge at room temperature for selected pressures.
		}\label{fig3}
	\end{center}
\end{figure}

\begin{figure*}[tb !]
\begin{center}
  \includegraphics[clip,width=1.99\columnwidth]{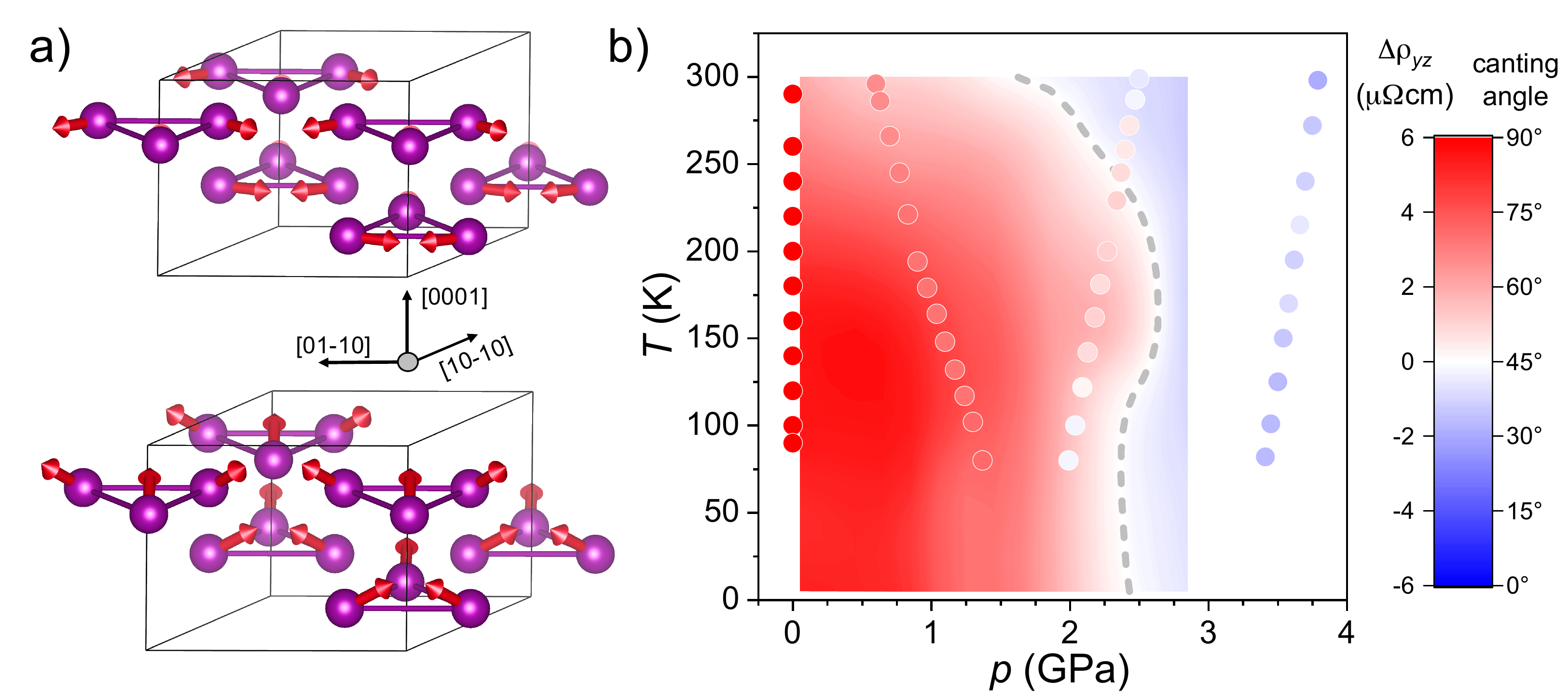}
  \caption{(a) The inverse triangular magnetic structure of Mn$_{3}$Ge at ambient pressure (top) and at $p=2.85$~GPa (bottom). (b) Pressure -- temperature \MN{}{color map of the AHE amplitude ($\Delta\rho_{\rm yz}$)} for the Mn$_{3}$Ge compound. The bullets represent the angle of the spin moments out of the \MN{}{$(0001)$}-plane extracted from previous neutron diffraction experiments \cite{Sukhanov,soh2020ground}. \MN{}{The gray dashed line indicates a vanishing AHE.}
  }\label{fig4}
  \end{center}
\end{figure*}

The most marked feature in Mn$_{3}$Ge is the appearance of the large AHE at room temperature. Therefore, controlling the amplitude of the AHE at room temperature is a key for any future application in devices. In Fig.\ \ref{fig3} we show the evolution of \MN{}{$\sigma_{yz}$} as a function of external pressure up to 2.85~GPa at 300~K. The saturation value of \MN{}{$\sigma_{yz}(H)$} can be continuously tuned by increasing pressure from positive to negative values, at positive fields and vice versa at negative fields. At 0.05~GPa, virtually ambient pressure, we observe a large $|\MN{}{\sigma^{sat}_{yz}}|$ of $40~{\rm \Omega^{-1}cm^{-1}}$ already for small applied fields. Increasing pressure leads to a dramatic reduction in the saturation value. \MN{}{$\sigma_{yz}$} becomes zero in the whole field range at $p=1.53$~GPa. More interestingly, for higher pressures \MN{}{$\sigma_{yz}$}  starts to change sign reaching a saturation signal $|\MN{}{\sigma^{sat}_{yz}}|=23~{\rm\Omega^{-1}cm^{-1}}$  at $p=2.85$~GPa.



At ambient pressure, the large anomalous Hall effect in Mn$_{3}$Ge was discovered before \cite{Nayak_2016,Naoki2016} and its appearance explained by a nonvanishing Berry curvature due to the noncollinear triangular antiferromagnetic spin structure \cite{Kubler_2014,Chen_PhysRevLett2014,Shindou_PhysRevLett.2001,Nakatsuji_2015,PhysRevB.95.075128,Zhang_2018}. Our experimental results show a direct correlation between application of hydrostatic pressure and the amplitude of the AHE \MN{suggesting that the Berry curvature can be tuned by external pressure.}{} As reported by Sukhanov \textit{et al.}\ in a neutron diffraction study \cite{Sukhanov}, application of external pressure does not substantially affect the $120^{\circ}$ triangular antiferromagnetic spin structures, but induces a dramatic modification of the out-of-plane moment of the spin triangle which can be quantified by the canting angle of the moment with respect to the \MN{}{$(0001)$}-plane. Upon increasing pressure the noncollinear triangular magnetic structure of Mn$_{3}$Ge changes gradually to a uniformly canted non-collinear triangular structure and successively to a collinear ferromagnetic structure above 5~GPa \cite{Sukhanov}, which is beyond the pressure range of our present study.

Figure \ref{fig4}(a) illustrates the arrangement of the Mn moments at ambient pressure and at 2.85~GPa, the highest pressure of our experiment, clearly indicating the out-of-plane tilting of the moments at 2.85~GPa.
The correlation of the pressure-induced changes in the AHE and the modifications in the magnetic structure of Mn$_{3}$Ge becomes evident in the plot shown in Fig.\ \ref{fig4}(b). In the temperature -- pressure \MN{}{color map} the continuous color code corresponds to the size of the jump in the saturation value of the Hall resistivity taken in negative and positive magnetic fields $\Delta \MN{}{\rho_{yz}}=2\times\MN{}{\rho^{sat}_{yz}}$. The  canting angle of the Mn moments with respect to the \MN{}{$(0001)$}-plane taken from Ref.\ \citealp{Sukhanov} is indicated by colorized bullets corresponding to the color scheme provided in Fig.\ \ref{fig4}b. As a result we find an astonishing correspondence between the jump size of the AHE signal and the canting angle in the whole $T-p$ plane. Thus, we can argue that the pressure-induced modifications of AHE are due to modifications of the magnetic arrangement of the Mn moments.

\MN{}{Application of pressure induces a gradual change of the magnetic structure of Mn$_3$Ge, with the moments starting to move toward the  $[0001]$-direction [see Fig.~\ref{fig4}(a)]. In our Hall setup, the current is applied along the $[0001]$-direction in parallel with the out-of-plane ferromagnetic moment. Therefore, the increasing out-of-plane ferromagnetic moment has no effect on the measured Hall signal. The magnetic configuration in the $(0001)$-plane stays essentially the same, but with a reduced size of the in plane components of the Mn moments.}

Beyond the sign change in the amplitude of \MN{}{$\sigma_{yz}$} at around $1.5$~GPa, the \MN{}{$\sigma_{yz}$} curves start to present a small magnetic hysteresis. The size is slightly increasing with pressure, reaching a `coercive field' of $\mu_0H_c\approx 0.25~T$.
The hysteresis appears due the special spin arrangement present in Mn$_{3}$Ge, which leads to a formation of antiferromagnetic domains. At ambient pressure the presence of a weak in-plane ferromagnetic moment ensures that the application of a small magnetic field in the \MN{}{$(0001)$}-plane is sufficient to prevent the formation of magnetic domains.
\MN{}{However, for higher pressures, with gradual rotation of the spin into [0001]-direction, the in-plane ferromagnetic moment is reduced, and therefore a larger magnetic field is needed for suppress the antiferromagnetic domains. \cite{Sukhanov}.}
We note that although we start to observe a hysteresis in the \MN{}{$\sigma_{yz}$} curves, \MN{}{$\sigma_{yz}$} saturates immediately and stays flat for the entire magnetic field range indicating that the in-plane ferromagnetic moment plays no role in the observed AHE except in defining the chirality of the spin triangular structure [see Fig.~\ref{fig1}(a)].

We have observed that at 300~K the AHE, $|\MN{}{\sigma^{sat}_{yz}}|\approx40~{\rm \Omega^{-1}cm^{-1}}$, is continuously suppressed as a function of increasing external pressure. The Hall signal vanishes completely in the full investigated magnetic field range at $p=1.53$~GPa. With further increasing pressure, the Hall signal changes sign and starts increasing again, reaching a maximum of  $|\MN{}{\sigma^{sat}_{yz}}|\approx23~{\rm \Omega^{-1}cm^{-1}}$ at $p=2.85$~GPa.
Considering previous powder neutron diffraction results, \MN{}{ these data seem to indicate that the observed changes in the AHE are directly connected to modifications in the magnetic configuration. We can then speculate that these are causing changes in the electronic band structure, which further lead to a sign change in the Berry phase. However, density-functional theory calculations  dealing with $d$-electrons in magnetic systems are challenging in the required precision and make this point an open question for further studies.}


In conclusion, we presented an investigation of the pressure evolution of the anomalous Hall effect in the chiral antiferromagnet Mn$_{3}$Ge. Our findings demonstrate that application of external pressure can play an efficient role in tuning the anomalous Hall effect in antiferromagnetic materials and offers a new route for further exploration of the Berry phase as the physical mechanism behind the AHE.


This work was financially supported by the ERC Advanced Grant No.\ 742068 "TOP-MAT". RDdR and LOK acknowledge financial support from the Brazilian agencies CNPq and FAPESP (Grants 2018/00823-0 and 2018/19015-1) and from the Max Planck Society under the auspices of the Max Planck Partner Group of the MPI for Chemical Physics of Solids, Dresden, Germany. SS acknowledges Science and Engineering Research Board of India for financial support through an Early Career Research Award (Grant No. ECR/2017/003186) and the award of a Ramanujan Fellowship (Grant No. SB/S2/RJN-015/2017)"


\bibliography{Mn3Ge-references}

\end{document}